%% file: LKGR.tex
\documentclass[sigconf]{acmart}
\usepackage{times}
\usepackage{soul}
\usepackage{url}
\usepackage{graphicx}
\usepackage{amsmath}
\usepackage{amsthm}
\usepackage{amsfonts}
\usepackage{booktabs}
\usepackage[linesnumbered,ruled,vlined]{algorithm2e}
\usepackage{color}
\usepackage{multirow}
\usepackage{enumitem}
\usepackage{float}
\usepackage{caption}
\usepackage{subfigure}
\usepackage{stfloats}
\usepackage{flushend}
\usepackage{indentfirst}
\usepackage{setspace}
\usepackage{balance}

\DeclareMathOperator\arcosh{arcosh}

\AtBeginDocument{%
  \providecommand\BibTeX{{%
    \normalfont B\kern-0.5em{\scshape i\kern-0.25em b}\kern-0.8em\TeX}}}

\copyrightyear{2022}
\acmYear{2022}
\setcopyright{acmcopyright}\acmConference[WSDM '22]{Proceedings of the Fifteenth ACM International Conference on Web Search and Data Mining}{February 21--25, 2022}{Tempe, AZ, USA}
\acmBooktitle{Proceedings of the Fifteenth ACM International Conference on Web Search and Data Mining (WSDM '22), February 21--25, 2022, Tempe, AZ, USA}
\acmPrice{15.00}
\acmDOI{10.1145/3488560.3498419}
\acmISBN{978-1-4503-9132-0/22/02}

\begin{document}
\fancyhead{}

\title{Modeling Scale-free Graphs with Hyperbolic Geometry for Knowledge-aware Recommendation}

\author{
Yankai Chen$^1$,
Menglin Yang$^1$,
Yingxue Zhang$^2$,\\
Mengchen Zhao$^2$,
Ziqiao Meng$^1$,
Jianye Hao$^2$,
Irwin King$^1$
}
 \affiliation{
  \institution{$^1$The Chinese University of Hong Kong, $^2$Huawei Noah’s Ark Lab}
  \city{}
  \country{}
}
\email{ {ykchen, mlyang, zqzhao, king}@cse.cuhk.edu.hk, {zhaomengchen, yingxue.zhang, haojianye}@huawei.com }

\input{abstract}

\ccsdesc[500]{Information systems~Recommender systems}

\keywords{recommender system, knowledge graph, hyperbolic geometric}

\maketitle

\input{intro}

\input{problem}

\input{method}

\input{exp}

\input{related}

\input{conclusion}



\bibliographystyle{ACM-Reference-Format}
\balance
\bibliography{ref}

\end{document}

%% file: abstract.tex
\begin{abstract}
Aiming to alleviate data sparsity and cold-start problems of traditional recommender systems, incorporating knowledge graphs (KGs) to supplement auxiliary information has recently gained considerable attention. 
Via unifying the KG with \textit{user-item interactions} into a tripartite graph, recent works explore the graph topologies to learn the low-dimensional representations of users and items with rich semantics.
These real-world tripartite graphs are usually \textit{scale-free}, however, the intrinsic hierarchical graph structures of which are underemphasized in existing works, consequently, leading to suboptimal recommendation performance.
To address this issue and provide more accurate recommendation, we propose a knowledge-aware recommendation method with \textit{Lorentz model} of the hyperbolic geometry, namely \textit{Lorentzian Knowledge-enhanced Graph convolutional networks for Recommendation (LKGR)}.
LKGR facilitates better modeling of scale-free tripartite graphs after the data unification. 
Specifically, we employ different information propagation strategies in the hyperbolic space to explicitly encode heterogeneous information from historical interactions and KGs.
Additionally, our proposed knowledge-aware attention mechanism enables the model to automatically measure the information contribution, producing the coherent information aggregation in the hyperbolic space.
Extensive experiments on three real-world benchmarks demonstrate that LKGR outperforms state-of-the-art methods by 3.6-15.3\% of Recall@20 on Top-K recommendation.

\end{abstract}

%% file: intro.tex
\section{Introduction}

\begin{figure}
\includegraphics[width=0.48\textwidth]{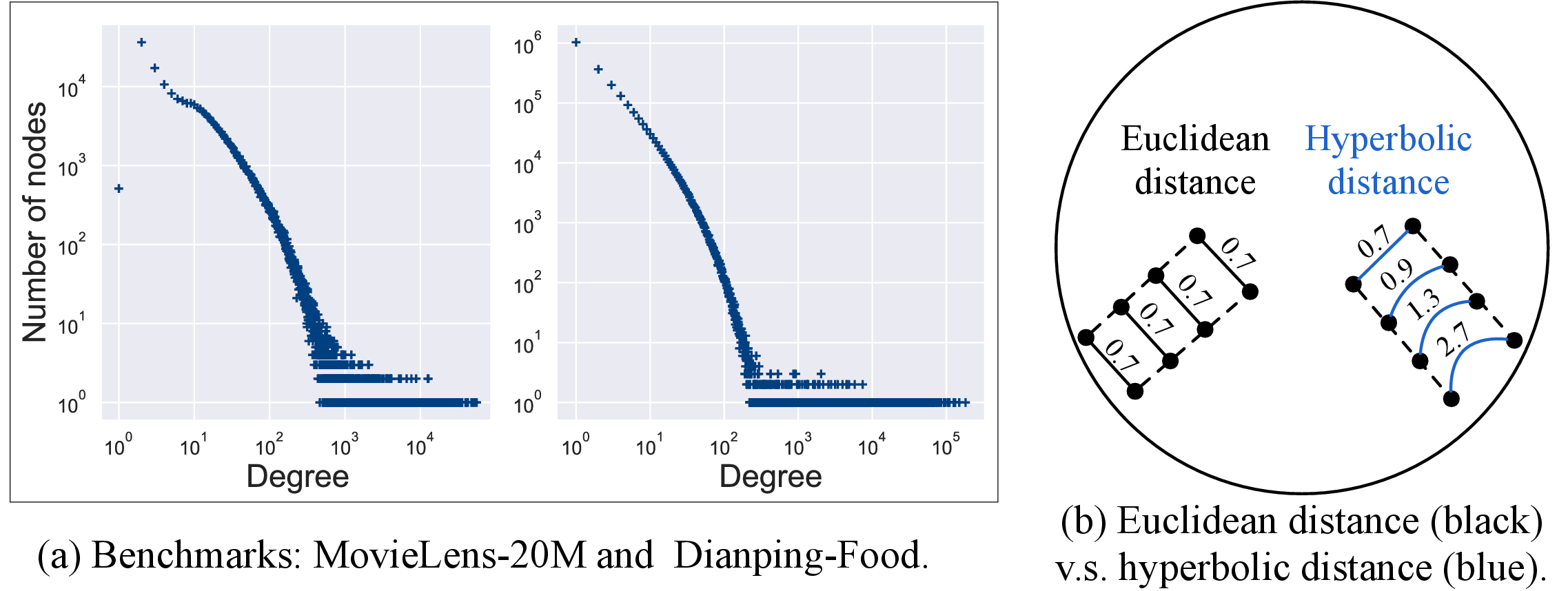}
\vspace{-0.2in}
\caption{(a) Degree distribution of two real-world benchmarks. (b) Distance comparison in Euclidean and hyperbolic spaces.}
\label{fig:distribution}
  \end{figure}

To alleviate the data sparsity and cold-start problems in traditional recommender systems~\cite{schein2002methods,herlocker2004evaluating,ma2007effective,koren2009matrix,volkovs2017dropoutnet,zhang2019star}, incorporating \textit{knowledge graphs} (KGs) into the recommender systems as side information has attracted growing attention in recent years~\cite{RippleNet,KGCN,KGAT,KGNNLS,CKAN}. 
A KG is a heterogeneous graph, where nodes represent \textit{entities} (i.e., products or items, as well as related attributes and properties) and edges represent mutual \textit{relations} between entities. 
Instead of relying on user-item historical records only, recommender systems (RS) extracting rich relational information in KGs can well compensate for the sparsity.
Recently, several works~\cite{KGCN,KGAT,CKAN,KGNNLS,RippleNet} develop~\textit{graph convolutional networks} (GCNs) in recommendation, thanks to their capability of modeling complex data dependency into the graph format and summarizing the semantic information behind the topology.

\textbf{Major motivation.} 
These GCN-based models for knowledge-aware recommendation~\cite{KGCN,KGAT,CKAN,KGNNLS,RippleNet} usually unify the historical \textit{user-item interactions} with \textit{KGs} into the tripartite graphs, as shown in Figure~\ref{fig:framework}(a). 
Then they learn the latent representations of users and items to estimate their matching probabilities in the Euclidean space.
However, after the data unification, these tripartite graphs present the \textit{scale-free} (hierarchical) graph characteristic, which is ignored by existing works~\cite{KGCN,KGAT,CKAN,KGNNLS,RippleNet}.
We analyze the real-world benchmarks that are widely studied in these works and show the degree distributions of two large datasets in Figure~\ref{fig:distribution}(a).
These two representative benchmarks are used to recommend movies\footnote{MovieLens-20M: \url{https://grouplens.org/datasets/movielens/}} and restaurants\footnote{Meituan-Dianping: \url{https://www.dianping.com/}} (details are in Section~\ref{sec:dataset}).
We observe that these graphs approximate the power-law distribution. 
According to Bourgain's theorem~\cite{linial1995geometry}, Euclidean space is however unable to obtain comparably low distortion for tree-like (power-law distributed) data~\cite{sala2018representation}.
Consequently, traditional graph embedding in the Euclidean space may not effectively capture the intrinsic hierarchical structures of these scale-free graphs, which leads to the high distortion of node embeddings and ultimately suppresses the recommendation performance~\cite{ravasz2003hierarchical,krioukov2010hyperbolic,chen2013hyperbolicity}.

To address this problem, we model these scale-free graphs with \textit{Lorentz model} of the hyperbolic geometry for knowledge-aware recommendation.
We propose an end-to-end model, namely \textit{Lorentzian Knowledge-enhanced Graph convolutional networks for Recommendation (LKGR)}. 
Generally, LKGR learns better representations for users, items, and KG entities for recommendation.
Concretely:
\begin{itemize}[leftmargin=*, topsep=2pt]
\item \textit{Firstly}, LKGR projects node representations onto the Lorentzian manifold, i.e., one specific Riemannian manifold.
It lives in the hyperbolic space, i.e., a continuous tree space with exponential volume growth property, producing less distortion when embedding scale-free data with intrinsic hierarchical structures~\cite{ravasz2003hierarchical,krioukov2010hyperbolic,chen2013hyperbolicity,yang2021discrete}. 
As shown in Figure~\ref{fig:distribution}(b), in the hyperbolic space, graph nodes that are closer to the graph center show a smaller distance, while nodes near the graph boundary present a larger distance. 
This exponentially-evolved distance measurement actually fits well with the tree-like graph structure, where nodes with small degrees can be viewed as leaves on the boundary, and nodes with large degrees are like roots located in the central positions.
On the contrary, Euclidean space embeddings are not position-sensitive and thus may not capture the latent graph hierarchical information.

\item \textit{Secondly}, at information propagation stage, unlike previous work such as KGCN~\cite{KGCN} and KGNN-LS~\cite{KGNNLS} mainly focusing on mining KGs, LKGR summarizes the \textit{interactive signals} from the user-item interactions and extracts \textit{informative knowledge} from KGs simultaneously.
Then we detach these two heterogeneous information from each other to update the node embeddings accordingly.
This is main because of the heterogeneity of graph nodes under the recommendation scenarios, which is also different from most previous works that conduct undifferentiated convolutional operations to all nodes.
These interactive signals are vital that directly reveal the user preferences and item targeting customers; along with the information extracted from KGs, LKGR further enriches the embeddings of users and items with diverse information components.

\item \textit{Thirdly}, we propose Lorentzian Knowledge-aware Attention Mechanism by considering the local graph structures on the Lorentzian manifold.
LKGR with our proposed attention mechanism can well weigh the relative importance of neighboring information and selectively propagate information on the associated manifold. 
In addition, LKGR is equipped with high-order information propagation techniques in the hyperbolic space, which enables it to be extensible for different recommendation benchmarks.  
\end{itemize}
 
LKGR learns the hierarchical structures of scale-free graphs with the hyperbolic geometry, and coherently summarizes interactive signals and knowledge associations into the low-dimensional embeddings.
We extensively evaluate LKGR in three real-world scenarios of book, movie, and restaurant recommendation, compared to recent state-of-the-art methods.
Experimental results not only prove the effectiveness of all proposed model components, but also demonstrate the superiority of LKGR over compelling baselines: we achieve the improved performance by 3.61-15.32\% of Recall metric for Top-20 recommendation.

\textbf{Organization.}
We define the problem in Section~\ref{sec:problem} and present the detailed methodology of LKGR model in Section~\ref{sec:method}. In Section~\ref{sec:exp}, we then report the experimental results. Finally, we review the related works in Section~\ref{sec:related} and conclude the paper in Section~\ref{sec:con}.

%% file: problem.tex
\section{Problem and Motivation}
\label{sec:problem}
\subsection{Problem Formulation}
A KG is formally defined as a set of the knowledge triplets: $\{(e_1,r,e_2)|e_1,e_2$ $\in$ $\mathcal{E}$, $r$ $\in$ $\mathcal{R}\}$, denoting that relation $r$ connects entity $e_1$ and $e_2$. $\mathcal{E}$ and $\mathcal{R}$ represent the set of entities and relations. The KG is used to provide external knowledge for items, e.g., (\textit{The Godfather, ActedBy, Marlon Brando}).
User-item interactions can be similarly represented as: $\{(u, r^*, i)|u$ $\in$ $\mathcal{U}$, $i$ $\in$ $\mathcal{I}\}$. $\mathcal{U}$ and $\mathcal{I}$ denote the set of users and items, respectively, and $r^*$ generalizes all interaction types, e,g., \textit{browse}, \textit{click}, or \textit{purchase}, as one relation between $u$ and $i$.
The user-item interaction matrix $\boldsymbol{Y} \in \mathcal{R}^{|\mathcal{U}| \times |\mathcal{I}|}$ is defined according to user-item interaction, where $y_{u,i}$ $=$ $1$ indicates there is an observed interaction, otherwise $y_{u,i}$ $=$ $0$.
Moreover, each item can be matched with an entity in the KG to elucidate alignments between items and entities~\cite{KGAT,CKAN}. 
Unifying user behaviors and item knowledge into the \textit{Unified Knowledge Graph} (UKG), which essentially is a tripartite graph and can be defined as $\mathcal{G}$ $=$ $\{(h,r,t)|h,t$ $\in$ $\mathcal{E}'$, $r$ $\in$ $\mathcal{R}'\}$ where $\mathcal{E}'$ $=$ $\mathcal{E}$ $\cup$ $\mathcal{U}$ and $\mathcal{R}'$ $=$ $\mathcal{R}$ $\cup$$\{r^*\}$. 
For example, the UKG shown in Figure~\ref{fig:framework}(a) unifies various relations as well as users, items, and entities in a graph for movie recommendation.

\textbf{Notations.} In this paper, we use the bold lowercase characters, bold uppercase characters and calligraphy characters to denote the vectors, matrices and sets, respectively. Non-bold characters are used to denote scalars or graph nodes.

\textbf{Task description.} Given the UKG, the recommendation task studied in this paper is to train a RS model estimating the probability $\hat{y}_{u,i}$ that user $u$ may adopt item $i$.

%% file: method.tex
\section{Proposed LKGR Approach}
\label{sec:method}

\begin{figure*}
\centering
\includegraphics[width=0.98\textwidth]{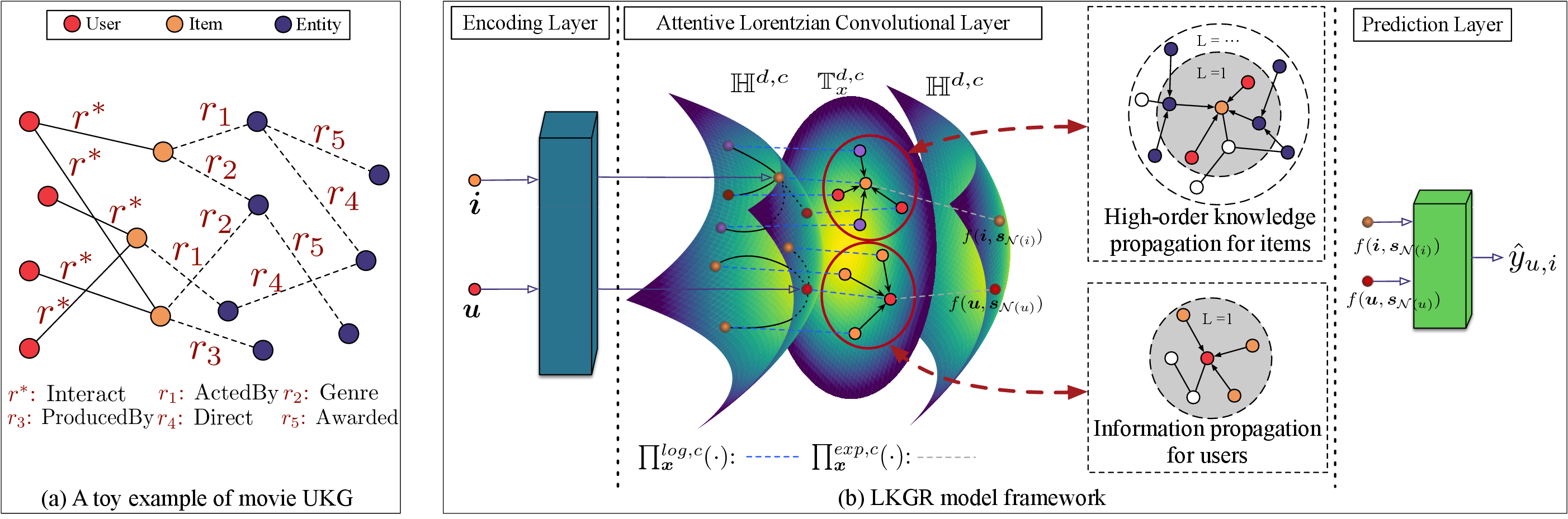}
\vspace{-0.1in}
\caption{(a) The tripartite graph modeling users, items and KG entities. (b) Illustration of the proposed LKGR model.}
\label{fig:framework}
  \end{figure*}

\subsection{Hyperbolic Geometry Preliminaries} 
Hyperbolic geometry is a non-Euclidean geometry with a constant negative curvature measuring how a geometric object deviates from a flat plane~\cite{robbin2011introduction}. 
We adopt the \textbf{Lorentz model}, i.e., one typical equivalent model that well describes hyperbolic geometry, for its unique simplicity and numerical stability~\cite{nickel2018learning,HGCN,law2019lorentzian,liu2019hyperbolic}. 

\textbf{Lorentzian manifold and tangent space.} Let $\left<.,.\right>_{\mathbb{H}}$: $\mathbb{R}^{d}$ $\times$ $\mathbb{R}^{d}$ $\rightarrow$ $\mathbb{R}$ represent the Lorentzian inner product in the hyperbolic space: 
\begin{equation}
\setlength\abovedisplayskip{2pt}
\setlength\belowdisplayskip{2pt}
 \left<\boldsymbol{x}, \boldsymbol{y} \right>_{\mathbb{H}} = -x_0y_0 + x_1y_1 + \dots + x_{d-1}y_{d-1}. \\
\end{equation}

For ease of presentation, the $d$-dimensional Lorentzian manifold is denoted by $\mathbb{H}^{d, c}$ with the negative curvature $-1/c$, and the Euclidean tangent space with $d$ dimensions centered at vector $\boldsymbol{x}$ $\in$ $\mathbb{H}^{d, c}$ is denoted by $\mathbb{T}_{\boldsymbol{x}}^{d, c}$. 
$\mathbb{T}_{\boldsymbol{x}}^{d, c}$ is a local, first-order approximation of the Lorentzian manifold at $\boldsymbol{x}$, which is useful to perform graph convolutional operations in the hyperbolic space~\cite{HGCN}.
%


\textbf{Exponential and logarithmic mappings.} Hyperbolic space and tangent space can be bridged by exponential and logarithmic mappings. 
Given $\boldsymbol{x}$, $\boldsymbol{y}$ $\in$ $\mathbb{H}^{d, c}$ ($\boldsymbol{x}$ $\neq$ $\boldsymbol{y}$), $\boldsymbol{z}$ $\in$ $\mathbb{T}_{\boldsymbol{x}}^{d, c}$ ($\boldsymbol{z}$ $\neq$ $\boldsymbol{0}$), 
the exponential mapping $\prod^{exp, c}_{\boldsymbol{x}}(\boldsymbol{z})$: $\mathbb{T}_{\boldsymbol{x}}^{d, c}$ $\rightarrow$ $\mathbb{H}^{d, c}$, maps $\boldsymbol{z}$ to the hyperbolic space; the reverse logarithmic mapping projects vectors back to the tangent space centered at $\boldsymbol{x}$, where $||\boldsymbol{z}||_{\mathbb{H}}$ $=$ $\sqrt{\left<\boldsymbol{z}, \boldsymbol{z}\right>_{\mathbb{H}}}$ is the norm of $\boldsymbol{z}$:
\begin{equation}
\setlength\abovedisplayskip{2pt}
\setlength\belowdisplayskip{2pt}
\begin{split}
& \begin{matrix}\prod\end{matrix}^{exp, c}_{\boldsymbol{x}}(\boldsymbol{z}) = \cosh(\frac{|| \boldsymbol{z}||_{{\mathbb{H}}}}{\sqrt{c}})\boldsymbol{x} + \sqrt{c}\sinh(\frac{||\boldsymbol{z}||_{\mathbb{H}}}{\sqrt{c}})\frac{\boldsymbol{z}}{||\boldsymbol{z}||_{{\mathbb{H}}}}, \\
& \begin{matrix}\prod\end{matrix}^{log, c}_{\boldsymbol{x}}(\boldsymbol{y}) = \sqrt{c}\arcosh(- \frac{\left<\boldsymbol{x}, \boldsymbol{y} \right>_{{\mathbb{H}}}}{c}) \cdot  \frac{\boldsymbol{y} + \frac{1}{c} \left<\boldsymbol{x}, \boldsymbol{y}\right>_{{\mathbb{H}}} \boldsymbol{x}}{|| \boldsymbol{y} + \frac{1}{c} \left<\boldsymbol{x}, \boldsymbol{y}\right>_{{\mathbb{H}}} \boldsymbol{x} ||_{{\mathbb{H}}}}.
\end{split}
\end{equation}

\subsection{LKGR Model Overview}
In the following content, we introduce the proposed Lorentzian Knowledge-enhanced Graph Convolutional Networks (LKGR) in detail. 
Figure~\ref{fig:framework}(b) illustrates the framework of the LKGR model. Generally, it consists of three components: 
\begin{itemize}[leftmargin=*, topsep=2pt]
\item \textbf{Encoding Layer.} 
In practice, input embeddings may live in the Euclidean space or hyperbolic space.
If they are in the Euclidean space, Encoding Layer first projects them to the hyperbolic space to make sure they are on the Lorentzian manifold.

\item \textbf{Attentive Lorentzian Convolutional Layer.} 
Aiming to accurately profile the latent representations of users and items, the Lorentzian Convolutional Layer respectively updates their embeddings based on their sampled ego-networks, e.g., Figures~\ref{fig:framework}(b) (unsampled nodes are colored white). 
In the hyperbolic space, users and items propagate \textit{interactive signals} back and forth, which actually simulates the collaborative filtering effect. 
To automatically learn the relative importance of information from the KG, we design \textit{Lorentzian Knowledge-aware Attention Mechanism}, which enables the selective and biased information aggregation in the hyperbolic space. Furthermore, by stacking multiple attentive Lorentzian Convolutional Layers, LKGR can explicitly explore the high-order KG subgraphs to further extract distant knowledge.

\item \textbf{Prediction Layer.} To achieve the efficient estimation in the matching stage, our Prediction Layer directly collects the learned representations of users and items and outputs the matching score, by conducting the inner product in the hyperbolic space. 
\end{itemize}

\subsection{Encoding Layer}
If input embeddings are in the Euclidean space, before inputting to the following layers, we first need to explicitly encode the Euclidean input onto the Lorentzian manifold. Let $\boldsymbol{x}_{\mathbb{E}}$ $\in$ $\mathbb{R}^{d-1}$ and $\boldsymbol{x}_{\mathbb{H}}$ $\in$ $\mathbb{H}^d$ denote Euclidean inputs and the transformed hyperbolic feature embedding, respectively. $\boldsymbol{x}_{\mathbb{H}}$ can be encoded as follows:
\begin{equation}
\setlength\abovedisplayskip{2pt}
\setlength\belowdisplayskip{2pt}
\resizebox{1\linewidth}{!}{$
    \displaystyle
\begin{aligned}
\boldsymbol{x}_{\mathbb{H}} & = \begin{matrix}\prod\end{matrix}^{exp, c}_{\boldsymbol{o}}([0, \boldsymbol{x}_{\mathbb{E}}]) 
= \left[\sqrt{c}\cosh(\frac{||\boldsymbol{x}_{\mathbb{E}}||_2}{\sqrt{c}}), \sqrt{c}\sinh(\frac{||\boldsymbol{x}_{\mathbb{E}}||_2}{\sqrt{c}}) \frac{\boldsymbol{x}_{\mathbb{E}}}{||\boldsymbol{x}_{\mathbb{E}}||_2}  \right],
\end{aligned}
$}
\end{equation}
where $(0, \boldsymbol{x}_{\mathbb{E}})$ is a $d$-dimensional vector in (Euclidean) tangent space and vector $\boldsymbol{o}$ $=$ $\{\sqrt{c}, 0, \dots, 0\}$ $\in$ $\mathbb{H}^{d, c}$ denotes the origin in $\mathbb{H}^{d, c}$.
$\boldsymbol{o}$ is used as a reference vector to perform tangent space operations. 
In the context of LKGR, we set the curvature $-1/c$ as a trainable variable, which dynamically measures how hierarchical the embedding space is~\cite{HGCN}.
Unless otherwise specified, we use $\boldsymbol{x}$ to denote $\boldsymbol{x_{\mathbb{H}}}$ in the following sections.

Encoding Layer enables our model to be compatible with upstream Euclidean inputs. 
To evaluate the holistic performance, e.g., recommendation accuracy, training time, of all proposed LKGR modules, in this paper, we initialize the node embeddings in the Euclidean space. 
Experimental details can be found in Section~\ref{sec:exp_setup}.

\subsection{Attentive Lorentzian Convolutional Layer}

\subsubsection{Lorentzian Knowledge-aware Attention} 
Our proposed attention mechanism considers the local structures of knowledge triplets on the Lorentzian manifold.
We first compute the attentive weight $\pi({h},{r},{t})$ of an edge between entity $h$ and entity $t$ connected by relation $r$:
\begin{equation}
\setlength\abovedisplayskip{2pt}
\setlength\belowdisplayskip{2pt}
\pi({h},{r},{t}) = \begin{matrix}\prod^{log, c}_{\boldsymbol{o}}(\boldsymbol{h})\end{matrix}^T \boldsymbol{W}_r \begin{matrix}\prod\nolimits^{log, c}_{\boldsymbol{o}}(\boldsymbol{t})\end{matrix},
\end{equation}
which is further normalized, denoted by $\hat{\pi}(h,r,t)$, across all edges connected with $h$ by adopting the softmax function:
\begin{equation}
\setlength\abovedisplayskip{2pt}
\setlength\belowdisplayskip{2pt}
\hat{\pi}(h,r,t) = \frac{\exp(\pi(h,r,t))}{\sum_{t' \in \mathcal{N}(h)} \exp(\pi(h,r,t'))}.
\end{equation}

Our attention mechanism depends on the node embedding $\boldsymbol{h}$, $\boldsymbol{t}$ and weight matrix $\boldsymbol{W}_r$, enabling LKGR to automatically measure different contributions of knowledge-based neighbors. Based on these learned weights, neighboring information can be selectively propagated and aggregated. 

\subsubsection{Lorentzian Information Propagation}
To propagate the neighbor information on the Lorentzian manifold, we compute the Lorentzian linear combination of neighborhoods for users and items, respectively. As shown in Figure~\ref{fig:framework}(a), since users only interact with items, the embedding of user $u$'s ego-network (i.e., $\mathcal{N}(u)$) representing $u$'s historical interactive information, is defined in Equation~\ref{eq:user}. Here $\hat{\pi}(u,r^*,i)$ is the normalized weight of edge $(u, r^*, i)$. 
\begin{equation}
\setlength\abovedisplayskip{2pt}
\setlength\belowdisplayskip{2pt}
\label{eq:user}
\boldsymbol{s}_{\mathcal{N}(u)} = \begin{matrix}\prod\end{matrix}^{exp, c}_{\boldsymbol{u}}\Big(\sum_{i\in \mathcal{N}(u)} \hat{\pi}(u,r^*,i) \begin{matrix}\prod\end{matrix}^{log, c}_{\boldsymbol{u}}(\boldsymbol{i}) \Big).
\end{equation}

Similarly, items connect to users and KG entities; therefore, item $i$ collectively receives the interactive signals from the user neighbors (i.e., $\mathcal{N}^{UI}(i)$), and knowledge from the KG side (i.e., $\mathcal{N}^{KG}(i)$). 
Let $u$ $\in$ $\mathcal{N}^{UI}(i)$ and $e$ $\in$ $\mathcal{N}^{KG}(i)$ denote a user and an entity connecting with item $i$ by relation $r^*$ and $r$. LKGR collectively summarizes the interactive information and knowledge backgrounds for item $i$ as:

\begin{equation}
\setlength\abovedisplayskip{2pt}
\setlength\belowdisplayskip{2pt}  
\label{eq:item}
\resizebox{1\linewidth}{!}{$
    \displaystyle  
\begin{aligned}
\boldsymbol{s}_{\mathcal{N}(i)} = \begin{matrix}\prod\end{matrix}^{exp, c}_{\boldsymbol{i}} 
\Big(\sum_{u\in \mathcal{N}^{UI}(i)} \hat{\pi}(i,r^*, u)  \begin{matrix}\prod\end{matrix}^{log, c}_{\boldsymbol{i}}(\boldsymbol{u}) 
+     \sum_{e\in \mathcal{N}^{KG}(i)} \hat{\pi}(i,r,e) \begin{matrix}\prod\end{matrix}^{log, c}_{\boldsymbol{i}}(\boldsymbol{e})\Big).
\end{aligned}
$}
\end{equation}

We implement a fixed-size random neighbor sampling instead of using full node neighbors, which is particularly useful for web-scale recommender systems~\cite{ying2018graph, graphSage}. 


\subsubsection{Lorentzian Neighbor Aggregation}
After obtaining the propagated neighbor information (i.e., $\boldsymbol{s}_{\mathcal{N}(u)}$ and $\boldsymbol{s}_{\mathcal{N}(i)}$), the next step is to perform Lorentzian neighbor aggregation and update the embeddings for users and items. 
Generally, for node embedding $\boldsymbol{x}$ and its neighbor representation $\boldsymbol{s}_{\mathcal{N}(x)}$, we use function $f(\boldsymbol{x},\boldsymbol{s}_{\mathcal{N}(x)})$: $\mathbb{H}^{d}$ $\times$ $\mathbb{H}^{d}$ $\rightarrow$ $\mathbb{H}^{d}$ to update the representation for node $x$, i.e., $\boldsymbol{x}$ $=$ $f(\boldsymbol{x}, \boldsymbol{s}_{\mathcal{N}(x)})$.
In this paper, we utilize three types of Lorentzian aggregators to implement $f(\cdot)$ as:
\begin{itemize}[leftmargin=*, topsep=2pt]
\item \textit{Sum Aggregator}~\cite{GCN} takes the summation of two inputs and conducts a nonlinear transformation, followed by a nonlinear activation on the Lorentzian manifold:
\begin{equation}
\setlength\abovedisplayskip{2pt}
\setlength\belowdisplayskip{2pt}
f_{sum} = \sigma^{\otimes^c} \Big( \boldsymbol{A} \odot^c ( \boldsymbol{x} \oplus^c \boldsymbol{s}_{\mathcal{N}(x)} ) \oplus^c \boldsymbol{b} \Big),
\end{equation}
where $\boldsymbol{A}$ and $\boldsymbol{b}$ are the trainable weights and bias defined in the associated tangent space. All Lorentzian basic operations e.g., $\odot^c$, will be introduced later.

\item \textit{Concatenate Aggregator}~\cite{graphSage} concatenates two vectors, followed by a nonlinear transformation and activation as:
\begin{equation}
\setlength\abovedisplayskip{2pt}
\setlength\belowdisplayskip{2pt}
f_{concat} = \sigma^{\otimes^c} \Big( \boldsymbol{A}\odot^c (\boldsymbol{x} \circledast^c \boldsymbol{s}_{\mathcal{N}{(x)}} ) \oplus^c \boldsymbol{b} \Big),
\end{equation}
where $\circledast^c$ denotes the operation:
\begin{equation}
\setlength\abovedisplayskip{2pt}
\setlength\belowdisplayskip{2pt}
\boldsymbol{a} \circledast^c \boldsymbol{b} = \begin{matrix}\prod\end{matrix}^{exp, c}_{o}\left(  \begin{matrix}\prod^{log, c}_{o}(\boldsymbol{a})\end{matrix} \Big|\Big| \begin{matrix}\prod^{log, c}_{o}(\boldsymbol{b})  \end{matrix} \right).
\end{equation}

\item \textit{Neighbor Aggregator}~\cite{GAT} directly updates the output representation with the input $\boldsymbol{s}_{\mathcal{N}(x)}$: 
\begin{equation}
\setlength\abovedisplayskip{2pt}
\setlength\belowdisplayskip{2pt}
f_{neighbor} = \sigma^{\otimes^c} \Big( \boldsymbol{A} \odot^c \boldsymbol{s}_{\mathcal{N}(x)} \oplus^c \boldsymbol{b} \Big).
\end{equation}
\end{itemize}

\textbf{Lorentzian linear transformation and activation.}
Given $\boldsymbol{A}$ and $\boldsymbol{b}$, Lorentzian linear transformation of the hyperbolic geometry can be well extended from the Euclidean geometry as~\cite{HNN,HGCN}:
\begin{equation}
\setlength\abovedisplayskip{2pt}
\setlength\belowdisplayskip{2pt}
\begin{split}
& \boldsymbol{A} \odot^c \boldsymbol{x} = \begin{matrix}\prod^{exp, c}_{\boldsymbol{o}}(\boldsymbol{A}  \prod^{log, c}_o(\boldsymbol{x}))\end{matrix}, \\
&\boldsymbol{x} \oplus^c \boldsymbol{b} = \begin{matrix}\prod^{exp, c}_{\boldsymbol{x}}\end{matrix} \Big(\boldsymbol{b} - \gamma\big(\begin{matrix}\prod\end{matrix}^{log, c}_{\boldsymbol{o}}(\boldsymbol{x}) + \begin{matrix}\prod\end{matrix}^{log, c}_{\boldsymbol{x}}(\boldsymbol{o}) \big) \Big), \\
\end{split}
\end{equation}
where $\gamma = \frac{ \left< \prod^{log, c}_{\boldsymbol{o}}(\boldsymbol{x}), \boldsymbol{b} \right>_{{\mathbb{H}}}  }{{c}\arcosh(- \frac{\left<\boldsymbol{o}, \boldsymbol{x} \right>_{{\mathbb{H}}}}{c})^2}$. The hyperbolic activation $\sigma$ is defined:
\begin{equation}
\setlength\abovedisplayskip{2pt}
\setlength\belowdisplayskip{2pt}
\sigma^{\otimes^c} = \begin{matrix}\prod^{exp, c}_{\boldsymbol{o}} \end{matrix}(\sigma(\begin{matrix}\prod\end{matrix}^{log, c}_{\boldsymbol{o}}(\boldsymbol{x}) )).
\end{equation}

\begin{algorithm}[h]
\small
\caption{LKGR algorithm}
\label{alg:LKGR}
\LinesNumbered  
\KwIn{UKG {\footnotesize $\mathcal{G}$}; trainable parameters {\footnotesize $\Theta$: $c$, $\{\boldsymbol{u}\}_{u\in\mathcal{U}}$, $\{\boldsymbol{i}\}_{i\in\mathcal{I}}$, $\{\boldsymbol{e}\}_{e\in\mathcal{E}}$, $\{\boldsymbol{W}_r\}_{r\in\mathcal{R}'}$, $\{\boldsymbol{A}_j, \boldsymbol{b}_j\}_{j=0}$}; hyper-parameters: {\footnotesize $d$, $L$, $\eta$, $\lambda$, $f(\cdot)$}. }
\KwOut{Prediction function $\mathcal{F}(u,i|\Theta, \mathcal{G})$} 
\While{\rm{LKGR not converge}}{
    \For{$(u,i) \in \mathcal{G}$ \rm{that} $y_{u,i}=1$}{
        $\mathcal{N}(u), \mathcal{N}^{UI}(i) \gets$ get sampled user-item neighbors for $u, i$; \\
        $\boldsymbol{s}_{\mathcal{N}(u)}, \boldsymbol{s}_{\mathcal{N}^{UI}(i)}$$\gets$propagate interactive information; \\
        $\boldsymbol{u} \gets f(\boldsymbol{u}, \boldsymbol{s}_{\mathcal{N}(u)})$; \\
        $\mathcal{N}^{KG}(i) \gets$ get sampled $L$-hops of KG neighbors for $i$; \\
        \For{$l = L, \cdots, 1$}{
          \For{$e \in$ \rm{($l$-$1$)-hop neighbor of } $i$ \text{in} $\mathcal{N}^{KG}(i)$}{
              $\mathcal{N}(e) \gets$ $e$'s entity neighbors in $\mathcal{N}^{KG}(i)^{(l)}$; \\
              $\boldsymbol{s}_{\mathcal{N}(e)} \gets$ propagate KG information; \\
              \If{$l = 1$}{
                  $\boldsymbol{s}_{\mathcal{N}(e)} \gets$ propagate interactive information and KG backgrounds to $e$; \\
              }
              $\boldsymbol{e} \gets f(\boldsymbol{e}, \boldsymbol{s}_{\mathcal{N}(e)})$; \\
          }
        }
      $\hat{y}_{u,i} \gets$ compute estimated matching score;\\
      $\mathcal{L} \gets $ compute loss and optimize LKGR model;\\ 
    }
}
\KwRet $\mathcal{F}$.\\
\end{algorithm}

\subsubsection{High-order Knowledge Extraction}
\label{sec:highorder}
To further explore the high-order information in KGs and propagate distant knowledge to items, as shown in Figure~\ref{fig:framework}(b), we can explore the multi-hop subgraphs and stack more propagation layers in LKGR accordingly.
Concretely, we first conduct $l$-hop neighbor sampling for item $i$ to reach its high-order subgraph $\mathcal{N}^{KG}(i)$ in KG, where we use $\mathcal{N}^{KG}(i)^{(l)}$ to represent $i$'s $l$-hop neighbors.
Then we propagate knowledge from the $l$-hop subgraph and iteratively aggregate to the centric node $i$.
For example, entity $e$ is the $l$-hop neighbor of item $i$ in KG, i.e., $e \in \mathcal{N}^{KG}(i)^{(l)}$. Then in the $l$-hop of propagation, we formulate $e$' neighbor representation by exploring $e$'s adjacent subgraph $\mathcal{N}^{KG}(i)^{(l+1)}$ as follows:
\begin{equation}
\setlength\abovedisplayskip{2pt}
\setlength\belowdisplayskip{2pt}
\label{eq:kg}
\boldsymbol{s}_{\mathcal{N}(e), \ e\in\mathcal{N}^{KG}(i)^{(l)}} = \begin{matrix}\prod\end{matrix}^{exp, c}_{\boldsymbol{e}}\Big(\sum_{e'\in \mathcal{N}^{KG}(i)^{(l+1)}} \hat{\pi}(e,r',e') \begin{matrix}\prod\end{matrix}^{log, c}_{\boldsymbol{e}}({\boldsymbol{e}')} \Big),
\end{equation}
where $\boldsymbol{e}'$, the embedding of entity $e'\in \mathcal{N}^{KG}(i)^{(l+1)}$, and coefficient $\hat{\pi}(e,r',e')$ are updated based on the previous step computation. 

Specifically, high-order propagation relies on the neighbor sampling to generate a multi-hop sub-graph where edges live in the consecutive hops. As illustrated by the pseudocodes in Algorithm~\ref{alg:LKGR}, the $l$-hop KG information can be iteratively propagated from $l=L$ to $l=1$ via message passing along these edges (lines 7-13). 
At the $1$-hop subgraph, the condensed KG information and user neighbor information collectively enriches $i$'s representation (lines 11-12). Please notice that $0$-hop neighbor of item $i$ in $\mathcal{N}^{KG}(i)$ is $i$ itself (line 8), so that if $l = 1$, $e = i$ and $\boldsymbol{s}_{\mathcal{N}(e)} = \boldsymbol{s}_{\mathcal{N}(i)}$ (lines 11-13).

\textbf{Time complexity analysis.} 
Let $Y$ denote the number of user-item interactions. 
$\alpha$ is the average time cost of numerical computation between Euclidean and hyperbolic spaces.
The training time cost per epoch (iteration) is $O\big(\alpha$$\cdot$$Y$$\cdot(${\small$|\mathcal{N}(u)|$}$+${\small$|\mathcal{N}^{UI}(i)|$}$+${\small$|\mathcal{N}^{KG}(i)|^L)\big)$}.
In this paper, for all benchmarks, the sampling size is no more than $8$.
Although the theoretical time complexity is exponential to $L$, in our work, $L$ $\leq$ $2$. 
This is because stacking too many propagation hops incurs performance detriment, the main cause of which lies in the well-known \textit{over-smoothing}~\cite{li2019deepgcns,li2018deeper} problem. 
As we will show in Section~\ref{sec:time}, compared to recent state-of-the-art models stacking limited hops ($L\leq 2$), LKGR is comparably efficient in practice.

\subsection{Prediction Layer and Model Optimization}
\subsubsection{Prediction Layer}
In traditional embedding-based matching models, inner product and L2 distance are widely adopted, mainly because this simple but effective interaction modeling enables accelerated computation on the online matching stage.
Therefore, we use the learned hyperbolic representations of users and items from the previous layer and take the inner product in the hyperbolic space to estimate their matching score:
\begin{equation}
\setlength\abovedisplayskip{2pt}
\setlength\belowdisplayskip{2pt}
\hat{y}_{u,i} = g(\boldsymbol{u}, \boldsymbol{i}) = \big(\begin{matrix}\prod\end{matrix}^{log, c}_o(\boldsymbol{u})\big)^T \begin{matrix}\prod\end{matrix}^{log, c}_o(\boldsymbol{i}).
\end{equation}

During the evaluation stage, items with top scores $\hat{y}_{u,i}$ are selected as recommended items to a given user $u$. 

\subsubsection{Model Optimization}
Let $\mathcal{S}^+_u$ denote the positive interacted item set of user $u$, i.e., $\hat{y}_{u,i} = 1$, and $\mathcal{S}^-_u$ represent the corresponding sampling set of negative items, i.e., $\hat{y}_{u,i} = 0$. 
To effectively optimize LKGR for training, in this paper, we set $|\mathcal{S}^+_u|$ $=$ $|\mathcal{S}^-_u|$. 
In each iteration of model training, $\mathcal{S}^+_u$ and $\mathcal{S}^-_u$ are updated accordingly. Finally, the loss function of LKGR is defined as follows:
\begin{equation}
\setlength\abovedisplayskip{2pt}
\setlength\belowdisplayskip{2pt}
\mathcal{L} = \sum_{u\in \mathcal{U}} \Big(
   \sum_{i\in \mathcal{S}^+_u} \mathcal{J}(y_{u,i}, \hat{y}_{u,i}) -  \sum_{i\in \mathcal{S}^-_u} \mathcal{J}(y_{u,i}, \hat{y}_{u,i}) \Big) + \lambda ||\Theta||_2^2.
\end{equation}
where $\mathcal{J}(\cdot)$ denotes the cross-entropy loss, $\Theta$ is the set of trainable model parameters and embeddings, and $||\Theta||_2^2$ is the $L$2-regularizer parameterized by $\lambda$.

%% file: exp.tex
\section{Experiments}
\label{sec:exp}
We evaluate LKGR model under the three real-world scenarios, with the aim of answering the following research questions:

\begin{itemize}[leftmargin=*, topsep=2pt]
\item \textbf{RQ1.} How does LKGR perform compared to state-of-the-art KG-enhanced recommendation models?
\item \textbf{RQ2.} How are the time efficiency of LKGR and baselines in model training?
\item \textbf{RQ3.} How do proposed model components and different hyper-parameter settings affect LKGR?
\end{itemize}

\subsection{Datasets}
\label{sec:dataset}

To evaluate the effectiveness of LKGR, we utilize the following three benchmarks for movie, book, and restaurant recommendations. In terms of the diversity in domain, size, and sparsity, all these three datasets are frequently evaluated in recent works~\cite{RippleNet,KGCN,KGNNLS,CKAN}.

Specifically, Book-Crossing\footnote{Book-Crossing: \url{http://www2.informatik.uni-freiburg.de/~cziegler/BX/}} (\textbf{Book}) is a dataset of book ratings (ranging from $0$ to $10$) extracted from the Book-Crossing community. MovieLens-20M (\textbf{Movie}) is a widely adopted benchmark for movie recommendation. It contains approximately 20 million explicit ratings (ranging from 1 to 5) on the MovieLens website. Dianping-Food (\textbf{Restaurant}) is a commercial dataset from Dianping.com that consists of over 10 million diverse interactions, e.g., clicking, saving, and purchasing, between about 2 million users and 1 thousand restaurants. The first two datasets are publicly accessible and the last one is contributed by Meituan-dianping Inc.~\cite{KGNNLS}. Dataset statistics are summarized in Table~\ref{tab:datasets}.

\begin{table}[t]
\centering
\caption{Three datasets used in this paper.}
\vspace{-0.1in}
\label{tab:datasets}
\setlength{\tabcolsep}{3.5mm}{
\begin{tabular}{|c | c  c  c|}
\toprule
                & Book    & Movie       & Restaurant\\
\midrule
\midrule
  \# users         & 17,860    & 138,159     & 2,298,698\\
  \# items         & 14,967    & 16,954      & 1,362\\
  \# interactions   & 139,746   & 13,501,622    & 23,416,418\\
\midrule[0.01pt]  
  \# entities     & 77,903    & 102,569     & 28,115\\
  \# relations      & 25      &   32      & 7\\
  \# KG triples     & 151,500   & 499,474     & 160,519 \\
 
\bottomrule
\end{tabular}}
\end{table}

\subsection{Baselines}
To demonstrate the effectiveness of the proposed model, we compare LKGR with three types of state-of-the-art recommendation methods: CF-based methods (BPRMF, NFM), regularization-based methods (CKE and KGAT), and propagation-based methods (RippleNet, KGCN, KGNN-LS, CKAN), as follows. 

\begin{itemize}[leftmargin=*, topsep=2pt]
\item \textbf{BPRMF}~\cite{BPRMF} is a representative CF-based method that performs matrix factorization with implicit feedback, optimized by the Bayesian Personalized ranking optimization criterion.
\item \textbf{NFM}~\cite{NFM} is a state-of-the-art neural factorization machine model for recommendation. 

\item \textbf{CKE}~\cite{CKE} is a classical regularization-based method. CKE learns semantic embeddings using TransR~\cite{TransR} with structural, textual and visual information to subsume matrix factorization under a unified Bayesian framework. 

\item \textbf{KGAT}~\cite{KGAT} is another representative regularization-based model that collectively refines user and item embeddings via an attentive embedding propagation layer. We use the pre-trained embeddings of users and items from BPRMF to initialize the model.

\item \textbf{RippleNet}~\cite{RippleNet} is a recent state-of-the-art propagation-based model. Aiming at enriching the users' representations, RippleNet uses a memory-like network to propagate users' preferences towards items by following paths in KGs.

\item \textbf{KGCN}~\cite{KGCN} is another state-of-the-art propagation-based method which extends spatial GCN approaches to the KG domain. By aggregating high-order neighbor information, both structure information and semantic information of the KG can be learned to capture users' potential long-distance interests.

\item \textbf{KGNN-LS}~\cite{KGNNLS} is a state-of-the-art propagation-based method that applies graph neural network architecture to KGs with label smoothness regularization for recommendation.

\item \textbf{CKAN}~\cite{CKAN} is the latest state-of-the-art propagation-based method which employs a heterogeneous propagation strategy to encode diverse information in KGs for recommendation.
\end{itemize}


\subsection{Experiment Setup}
\label{sec:exp_setup}

To evaluate LKGR, we randomly divide each dataset 5 times into training, evaluation, and test sets with the ratio of 6:2:2. In the evaluation of Top-K recommendation task, we use the model learned from the training set to rank K items for each user in the test set with the highest predicted scores $\hat{y}_{i,j}$. We choose two widely-used evaluation protocols, i.e., Recall@K and NDCG@K. 
We optimize all models with Adam optimizer~\cite{Adam} and adopt the default Xavier initializer~\cite{Xavier} to initialize the model parameters.

We implement the LKGR model under Python 3.7 and Pytorch 1.4.0. 
The experiments are run on a Linux machine with 1 GTX-2080TI GPU, 4 Intel Xeon CPU (Skylake, IBRS), 16 GB of RAM. 
For all the baselines, we follow the official hyper-parameter settings from original papers or as default in corresponding codes. For methods lacking recommended settings, we apply a grid search for hyper-parameters. The learning rate, denoted by $\eta$, is tuned within \{$10^{-3}, 5\times10^{-2}, 10^{-2}, 5\times10^{-1}$\} and the coefficient of $L2$ normalization is tuned among \{$10^{-6}, 10^{-5}, 10^{-4}, 10^{-3}$\}. The embedding size is searched in \{$8, 16, 32, 64, 128$\}.

\begin{table*}[t]
  \caption{Average results of Top@20 recommendation task. Underline indicates the second-best model performance. Bold denotes the empirical improvements against second-best (underline) models, and $*$ denotes scenarios where a Wilcoxon signed-rank test indicates a statistically significant improvement under 95\% confidence level between our model and second-best models.}
  \vspace{-0.1in}
  \label{tab:top20}
  \setlength{\tabcolsep}{3mm}{
  \begin{tabular}{|c|c c|c c|c c|}
    \toprule
    \multirow{2}*{Model}  & \multicolumn{2}{c|}{Book-Crossing} & \multicolumn{2}{c|}{MovieLens-20M} & \multicolumn{2}{c|}{Dianping-Food} \\
               ~  & Recall@20(\%) & NDCG@20(\%) & Recall@20(\%) & NDCG@20(\%) & Recall@20(\%) & NDCG@20(\%) \\
    \midrule
    \midrule
    BPRMF     &   4.67 (8.7e-3) & 2.80 (4.3e-3) & 20.48 (1.6e-2) & 15.77 (9.1e-3) & 19.90  (3.0e-2) & 10.79  (2.0e-2)\\
    
    NFM     &   3.93 (2.2e-2) & 2.17 (1.5e-2) & 19.79 (3.3e-2) & 14.28 (1.1e-2) & 23.85 (3.9e-2) & \underline{12.48} (3.0e-2)\\
    
    CKE     &  4.38  (9.6e-3) & 2.24 (4.2e-2) & 21.52 (1.2e-2) & 15.73 (1.3e-2) & 22.24 (3.1e-2) & 12.09 (1.5e-2)\\
    
    RippleNet   &  7.12 (2.1e-2) & 5.09 (1.7e-2) & 13.74 (2.6e-2) & \ \ 9.77 (1.7e-2) & 21.20 (4.1e-2) & 10.99 (2.0e-2)\\
    
    KGNN-LS   & \underline{8.51} (2.2e-2) & \underline{6.06} (1.7e-2) & 20.20 (1.0e-2) & 15.49 (1.3e-2) & 15.52 (4.9e-2) &\ \ 7.92 (2.9e-2)\\
    
    KGCN    & 7.85 (2.9e-2) & 5.93 (2.3e-2) & 19.24 (3.2e-2) & 13.87 (1.6e-2) & 19.03 (3.0e-2) & \ \ 9.34 (1.5e-2)\\
    
    KGAT    & 5.34  (6.1e-3) & 3.01 (7.9e-3) & \underline{21.80} (7.7e-3) & \underline{16.81} (1.1e-2) & 15.57 (2.4e-2) & \ \ 7.67 (1.7e-2)\\
    
    CKAN      & 6.19 (1.1e-2) & 3.47 (5.3e-3)  & 17.48 (1.7e-2) & 12.48 (1.4e-2) & \underline{24.10} (3.9e-2) & \textbf{13.33} (2.0e-2)  \\
    
    \midrule
    \midrule
    \textbf{LKGR} & \textbf{9.48$^*$} (2.3e-2) & \textbf{6.50$^*$} (1.6e-2) & \textbf{25.14$^*$} (4.1e-2) & \textbf{18.34$^*$}  (4.7e-2) & \textbf{24.97$^*$}  (4.4e-2) & 10.45 (1.9e-2)\\
    \% Improv.       & 11.40\%     & 7.26\%       & 15.32\%     & 9.10\%   & 3.61\%   & N/A   \\
    \bottomrule
  \end{tabular}}
\end{table*}

\begin{figure*}[t]
\centering
\includegraphics[width=1\textwidth]{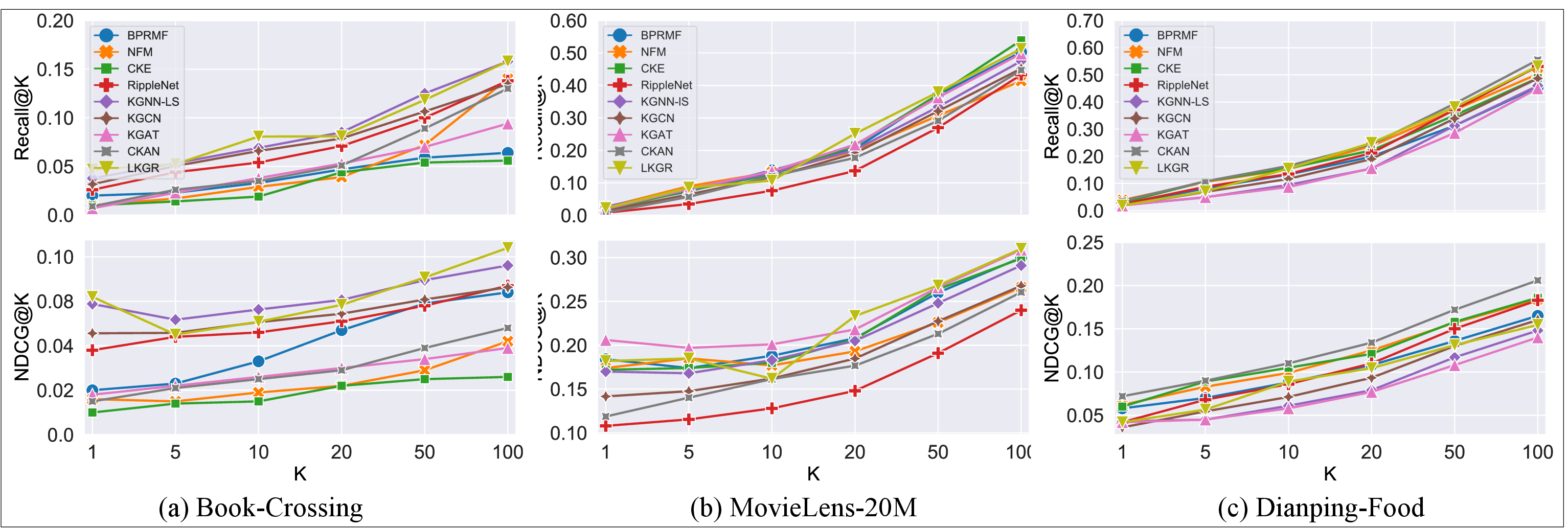}
\vspace{-0.2in}
\caption{Average results of Recall@K and NDCG@K in Top-K Recommendation.}
\label{fig:topk}
\end{figure*}

\subsection{Performance Comparison (RQ1)}

\subsubsection{Top-K recommendation}
We evaluate LKGR on Top-K recommendation task by varying K in \{$1, 5, 10, 20, 50, 100$\}. Table~\ref{tab:top20} reports Top-20 recommendation results for detailed comparison and analysis and Figure~\ref{fig:topk} contains the complete performance curves of LKGR and baselines. From these results, we can observe that:

\begin{itemize}[leftmargin=*, topsep=2pt]
\item \textit{The results of Top-20 recommendation demonstrate the improvements of LKGR are statistically stable and significant.}
We report the details of Top-20 recommendation of all models in Table~\ref{tab:top20}. For example, we can observe that LKGR significantly outperforms all baselines by 11.40\% (9.48$\rightarrow$8.51), 15.32\% (25.14$\rightarrow$21.80) and 3.61\% (24.97$\rightarrow$24.10) of Recall metric for Top-20 recommendation on three datasets. As for the NDCG metric, LKGR achieves 7.26\% (6.50$\rightarrow$6.06) and 9.10\% (18.34$\rightarrow$16.81) improvement of NDCG@20 on Book and Movie datasets but does not perform the best on Restaurant. This shows that LKGR can make good recalling of Top-20 candidates from the item corpus but may not well capture the ranking within for this specific dataset, leaving the space of further optimization.
The reported standard deviations of LKGR are also comparable with all baselines, showing the stability of our proposed model. 
Furthermore, we conduct \textit{Wilcoxon signed-rank tests} to verify that the achieved performance improvements are statistically significant over the second-best recommendation model under the 95\% confidence level. 

\item \textit{As K increases, LKGR consistently performs well on three benchmarks.}
Results in Figure~\ref{fig:topk} demonstrate the effectiveness of LKGR on improving the performance of the ranking recommendation task, i.e., Top-K recommendation. 
This shows that the hyperbolic geometry does well capture the hierarchical properties of graphs, meaning that it can preserve the order of users' preferences towards the items, which benefits the ranking recommendation task.
Furthermore, our discrete information propagation and aggregation strategies for users and items on the Lorentzian manifold are effective. We will conduct a more comprehensive ablation study in the later section, analyzing the contribution of all LKGR module components to the recommendation performance. 
\end{itemize}

\subsection{Time Efficiency Comparison (RQ2)}
\label{sec:time}

\begin{figure}[htp]
\small 
\begin{tabular}{c c}  
\hspace{-0.4in}
\begin{minipage}{0.2\textwidth}
\setlength{\tabcolsep}{0.5mm}{
  \begin{tabular}{|c | c  c  c|}
  \toprule
  Model     & BK  & MV & RT \\
  \midrule
  \midrule
    BPRMF     & { 4.58}    & { 109.21}  & { 134.97}  \\
    NFM       & { 26.50}   & { 92.52}   & { 313.87}  \\
    CKE       & { 10.15}   & { 82.83}   & { 98.61}   \\
    RippleNet & { 12.46}   & { 1,159.32} & { 2,343.93} \\
    KGNN-LS   & { 1.62}    & { 36.51}   & { 81.17}   \\
    KGCN      & { 4.47}    & { 15.27}   & { 58.80}   \\
    KGAT      & { 61.88}   & { 333.86}  & { 2,254.83}  \\
    CKAN      & { 3.36}    & { 439.23}  & { 529.21}   \\
  \midrule
  LKGR  & { 43.26}   & {250.88}  & {337.10}  \\
  \bottomrule
  \end{tabular}}
  \end{minipage}
  &
  \begin{minipage}{0.3\textwidth}
  \includegraphics[width=4.1cm]{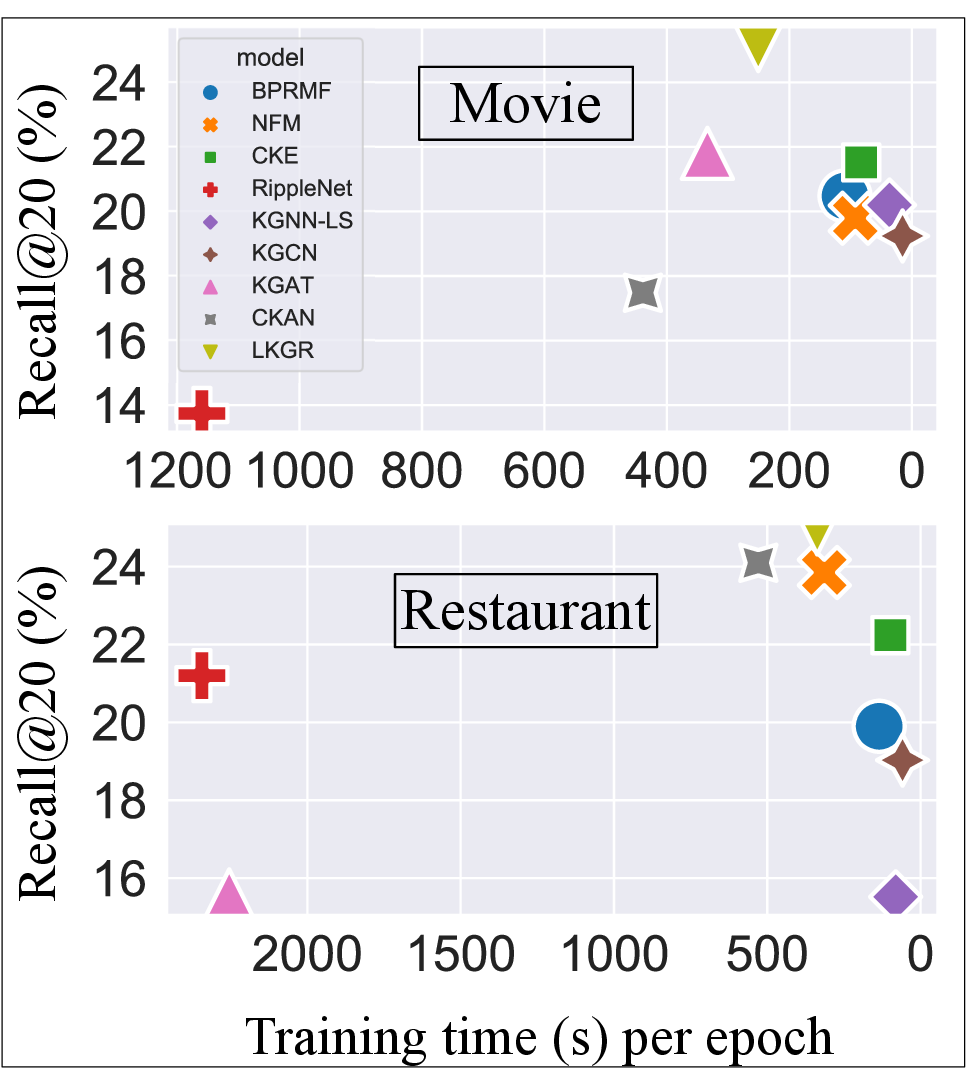}
  \end{minipage}
  \\
  \small (a) Training time cost per epoch (s).
  &
  \hspace{-0.5in}
  \small (b) Efficiency v.s. accuracy
\end{tabular}
\vspace{-0.1in}
\caption{Model comparison on efficiency and accuracy.}
\label{fig:time}
\end{figure}

In this section, we study how time efficient our LKGR and baselines are. 
All methods are run on the same aforementioned running environment, and we use the default hyper-parameters that are reported in papers or official codes.
We use BK, MV, and RT to denote Book, Movie, Restaurant, and report the results in Figure~\ref{fig:time}.

We can observe that: (1) as shown in Figure~\ref{fig:time}(a), compared to CF-based methods, i.e., BPRMF and NFM, LKGR requires more time for model training as it needs to explore the KG for information propagation; (2) compared to the remaining KG-based recommendation models, LKGR requires the analogous time cost of model training, which may dispel concerns of large computation overhead in the hyperbolic space. 
(3) Figure~\ref{fig:time}(b) visualizes the overall performances in terms of efficiency and accuracy among all models. 
As the upper-right corner of the figure means the ideal optimal performance, LKGR makes an excellent trade-off w.r.t efficiency and accuracy, especially on Movie and Restaurant datasets.

\subsection{Ablation Study of LKGR (RQ3.A)}
In this section, we first conduct a comprehensive ablation study to evaluate the effectiveness of all model components. Then we discuss how key hyper-parameters affect the recommendation performance.
We use R@20 and N@20 to denote recall@20 and NGCG@20.

\begin{table}[ht]
\centering
\caption{Ablation study on Top-K recommendation (\%). }
\vspace{-0.1in}
\label{tab:ablation}
\hspace{-0.15in}
\setlength{\tabcolsep}{0.3mm}{
\begin{tabular}{|c| c | c | c | c |c|}
\toprule
 Dataset      &   w/o IS &  w/o KE &  w/o HG  & w/o LKA & LKGR \\
\midrule
\midrule

        {\small BK-R@20}   & {\small \ \,4.75}{\,{\scriptsize(-49.89\%)}}    
                                & {\small \ \,5.79}{\,{\scriptsize(-38.92\%)}}   
                                & {\small \ \,8.11}{\,{\scriptsize(-14.45\%)}}
                                & {\small \ \,8.35}{\,{\scriptsize(-11.92\%)}}  
                                &{\small\ \,\textbf{9.48}} \\
        {\small BK-N@20}   & {\small \ \,3.81}{\,{\scriptsize(-41.38\%)}}   
                                & {\small \ \,4.22}{\,{\scriptsize(-35.08\%)}} 
                                & {\small \ \,6.21}{\ \,{\scriptsize(-4.46\%)}}  
                                & {\small \ \,6.43}{\ \,{\scriptsize(-1.08\%)}} 
                                &{\small\ \,\textbf{6.50}} \\   
 \midrule
        {\small MV-R@20}   & {\small 12.45}{\,{\scriptsize(-50.48\%)}}       
                                & {\small 11.73}{\,{\scriptsize(-53.34\%)}}      
                                & {\small 22.19}{\,{\scriptsize(-11.73\%)}}     
                                & {\small 23.94}{\ \,{\scriptsize(-4.77\%)}}    
                                &{\small\textbf{25.14}} \\
        {\small MV-N@20}   & {\small \ \,8.76}{\,{\scriptsize(-52.23\%)}}       
                                & {\small \ \,7.92}{\,{\scriptsize(-56.82\%)}}       
                                & {\small 15.47}{\,{\scriptsize(-15.65\%)}}   
                                & {\small 17.03}{\ \,{\scriptsize(-7.14\%)}}   
                                &{\small\textbf{18.34}} \\   
 \midrule
        {\small RT-R@20}   & {\small 11.22}{\,{\scriptsize(-55.07\%)}}     
                                & {\small \ \,9.86}{\,{\scriptsize(-60.51\%)}}    
                                & {\small 21.47}{\,{\scriptsize(-14.02\%)}}   
                                & {\small 23.17}{\ \,{\scriptsize(-7.21\%)}}     
                                &{\small\textbf{24.97}} \\
        {\small RT-N@20}   & {\small \ \,6.54}{\,{\scriptsize(-37.42\%)}}   
                                & {\small\ \,4.74}{\,{\scriptsize(-54.64\%)}}     
                                & {\small 10.41}{\ \,{\scriptsize(-0.38\%)}}    
                                & {\small 11.03}{\ \,{\scriptsize( 5.55\%)}} 
                                &{\small {10.45}} \\   
\bottomrule
\end{tabular}}
\end{table}

 \textbf{Impact of interactive signal propagation.}
To investigate the impact of information propagation and aggregation of interactive signal in LKGR, we set a variant model, denoted by LKGR$_{\rm w/o \ IS}$, which disables the information passing between users and items in our proposed Equations~(\ref{eq:user}) and~(\ref{eq:item}), so that items receive information only from KG entities. 
Based on the results reported in Table~\ref{tab:ablation}, without information propagation between users and items, the performance of Top-K recommendation drops dramatically, which demonstrates that learning historical interactive information for users and items is essential for the performance improvement.  

 \textbf{Impact of knowledge extraction.}
Similarly, we set another variant LKGR$_{\rm w/o \ KE}$ to study the influence of knowledge extraction, by cutting the KG information propagation to items (formulated in Equations~(\ref{eq:item}) and~(\ref{eq:kg})). As shown in Table~\ref{tab:ablation}, purely relying on the user-item interaction modeling is not enough to boost performance; thus integrating KGs in recommendation is also very important.

\textbf{Impact of hyperbolic geometry.}
We verify the impact of the hyperbolic geometry by replacing all hyperbolic operations into Euclidean space and retaining the computation logic of LKGR. We denote the variant as LKGR$_{\rm w/o \ HG}$. As shown in Table~\ref{tab:ablation}, even with adequate information propagation from user-item interactions and KGs, directly modeling these two data sources in Euclidean space brings about large performance decay on Top-K recommendation task. 
For example, by removing the hyperbolic geometry, LKGR$_{\rm w/o \ HG}$ produces worse listwise ranking results on all three benchmarks: 14.45\% (8.11$\rightarrow$9.48), 11.73\% (22.19$\rightarrow$25.14), and 14.02\% (21.47$\rightarrow$24.97) of Recall for Top-20 recommendation, respectively. 
To conclude, with all other essential components, embedding graph nodes on the Lorentzian manifold is substantial to LKGR, and modeling the scale-free graphs in the hyperbolic space can learn better representation for recommendation.

\textbf{Impact of Lorentzian Knowledge-aware Attention mechanism.}
We also examine the impact of our proposed attention mechanism. For comparison, we use a variant that set $\pi(h,r,t) = 1$ to make equal contribution for computing $\boldsymbol{s}_{\mathcal{N}(\cdot)}$, termed by LKGR$_{\rm w/o \ LKA}$. As shown in Table~\ref{tab:ablation}, disabling our attention mechanism leads to the decreased Recall@20 by 11.92\% (8.35$\rightarrow$9.48), 4.77\% (23.94$\rightarrow$25.14), and 7.21\% (23.17$\rightarrow$24.97). 
This proves that Lorentzian Knowledge-aware Attention mechanism is also effective, as it adaptively measures the importance weights for different nodes on the Lorentzian manifold and enables more coherent embedding aggregation.

\subsection{Effect of Key Hyper-parameters (RQ3.B)}
Due to the space limit, we only report the effect of some key hyper-parameters on the model performance.

\textbf{Effect of Lorentzian aggregators.}
To explore the influence of aggregating neighbor information on the Lorentzian manifold, we conduct experiments over different selections of Lorentzian aggregators $f(\cdot)$. Based on the Recall and NDCG metrics, we observe that in Table~\ref{tab:agg} \textit{sum aggregator}, i.e., $f_{sum}(\cdot)$, is superior on datasets of Book and Movie. For the Restaurant dataset, $f_{concat}(\cdot)$ is much more suitable. This is because sum and concat aggregators are capable of retaining external neighbor information as well as the internal information when comparing to $f_{neighbor}(\cdot)$.

\begin{table}[htbp]
\centering
\caption{Top-K recommendation (\%) w.r.t different $f$.}
\vspace{-0.1in}
\label{tab:agg}
\setlength{\tabcolsep}{2.1mm}{
\begin{tabular}{|c | c c | c c |c c|}
\toprule
\multirow{2}*{$f$}  & \multicolumn{2}{c|}{$f_{sum}$} & \multicolumn{2}{c|}{$f_{concat}$} & \multicolumn{2}{c|}{$f_{neighbor}$} \\
               ~  & R@20 & N@20 & R@20 & N@20 & R@20 & N@20 \\
\midrule
\midrule
  BK           & \textbf{9.48} & \textbf{6.50} & 8.19  & 6.06 & 8.10 & 5.84 \\
  MV          & \textbf{25.14} & \textbf{18.34}  & 21.42 & 16.33 & 20.87 & 15.23 \\
  RT        & 22.55 & 9.83 & \textbf{24.97} & \textbf{10.45} &  21.48   & 9.29 \\
\bottomrule
\end{tabular}}
\end{table}

\textbf{Effect of KG propagation depth $L$.}
We verify how the propagation depth affects the performance by varying $L$ from 0 to 3.
Depth 0 means no local information aggregation from KG. As shown in Table~\ref{tab:layer}, we notice that LKGR achieves the best performance when $L$ is 1, 2, and 1 for Book, Movie, and Restaurant. 
In addition to the over-smoothing issue that we explain in Section~\ref{sec:highorder}, another possible reason is: when long-distance propagation introduces distant knowledge, it may also bring about irrelevant information, especially when the dataset is large and dense. Thus, preserving an appropriate depth in the high-order information propagation enables maximized performance over different recommendation benchmarks adaptively.

\begin{table}[htbp]
\centering
\caption{Top-K recommendation (\%) w.r.t different $L$.}
\vspace{-0.1in}
\label{tab:layer}
\setlength{\tabcolsep}{0.7mm}{
\begin{tabular}{|c | c c | c c | c c | c c |}
\toprule
\multirow{2}*{$L$}  & \multicolumn{2}{c|}{$L=0$} & \multicolumn{2}{c|}{$L=1$} & \multicolumn{2}{c|}{$L=2$} & \multicolumn{2}{c|}{$L=3$} \\
               ~  & R@20 & N@20 & R@20 & N@20 & R@20 & N@20 & R@20 & N@20\\
\midrule
\midrule
  BK          & 5.79 & 4.22  & \textbf{9.48} & \textbf{6.50}  & 8.16 & 5.73   & 7.43  & 5.28 \\
  MV          & 22.19 & 15.47 & 20.34 & 15.06 & \textbf{25.14} & \textbf{18.34}  & 21.48 & 15.17 \\
  RT          & 9.86 & 4.74 & \textbf{24.97} & \textbf{10.45} &  22.12 & 9.47 & 21.67 & 9.02 \\
\bottomrule
\end{tabular}}
\end{table}

%% file: related.tex
\section{Related Works}
\label{sec:related}

With the rapid development of information networks, studying the ubiquitous graph data has aroused various interests in both industry and research communities~\cite{fang2017effective,IDX,yang2020featurenorm,zhang2019,defferrard2016}. 
Among these research topics, KG-enhanced recommender systems receive much attention recently as they can alleviate data sparsity and cold-start problems for better recommendation.
These methods can be generally categorized into three branches: (1) path-based, (2) regularization-based, and (3) propagation-based.

(1) In \textbf{path-based methods}, high level meta-paths are extracted from KGs and then input into the predictive model. 
Such meta-paths are usually manually selected or generated, which require intensive input of domain knowledge and labor resources~\cite{hu2018leveraging,MAGNN20}. 
Furthermore, it is difficult to optimize the path retrieval for large and complex graphs, while the selected paths have a great impact on the final performance.
Thus we exclude path-based methods for model comparison.
(2) In \textbf{regularization-based methods}, additional loss terms are devised to capture the KG structure and regularize the model training. 
For example, KGAT~\cite{KGAT} merges the two tasks of recommendation and KG completion to jointly train the model. 
HyperKnow~\cite{HyperKnow} further embeds the joint training in Poincar\'e Ball with an adaptive regularization mechanism to balance two loss terms.
However, one deficiency is that these methods usually rely on the traditional knowledge graph embedding methods to complete the KG training, high-order semantic information in the KG and user-item interactions is not explicitly modeled, which may result in suboptimal representation learning for users and items.
%
(3) \textbf{Propagation-based methods}, performing iterative message passing with the guidance of the topological structure in the KGs~\cite{RippleNet,KGNNLS,KGCN,CKAN}, has attracted much attention recently. With the auxiliary information passed along links in the KGs, the representations of users and items can be refined to provide more accurate recommendation service. 
However, all these methods assume the learning process in the low-dimensional Euclidean space; nevertheless, because of the intrinsic hierarchical structure of KGs, whether the Euclidean space is appropriate for all kinds of scenarios is still an open question.
Conceptually, LKGR is inspired by the graph-based message passing mechanism and performs graph convolutional operations in the hyperbolic space. 

%% file: conclusion.tex
\section{Conclusion and Future Works}
\label{sec:con}
In this paper, we propose a KG-enhanced recommendation model, namely LKGR, which learns the embeddings of users and items as well as the KG entities in the hyperbolic space.
We propose a knowledge-aware attention mechanism on the Lorentzian manifold to discriminate the contribution of graph node informativeness, which is followed by multi-layer aggregation for high-order information propagation. 
The experimental results over three real-world datasets not only validate the performance improvements of LKGR over recent state-of-the-art solutions, but also demonstrate the effectiveness of all proposed model components.

As for the future works, there are two potential directions: (1) since all existing works unify the user-item interactions and the KG into a static graph, while in practice, users and items usually contain temporal and dynamic interactions. How to simultaneously learn the graph structure and temporal information in a unified framework is a good direction to work on. 
(2) It is worth exploring other application scenarios with hyperbolic geometry for performance improvement, e.g., information retrieval~\cite{zhang2019doc2hash} and language processing~\cite{gao2020discern,li2020unsupervised}, such that hyperbolic modeling can well learn the embedding of interrelated data with less distortion.